\documentclass[seceq]{ptptex}

\usepackage{graphicx} 

\newcommand{\EV}{\vec{E}}
\newcommand{\BV}{\vec{B}}

\title{
Some Properties of
an Axisymmetric Pulsar  Magnetosphere 
Constructed by Numerical Calculation
}

\author{%
Jun  \textsc{Ogura} 
and
Yasufumi \textsc{Kojima}
}

\inst{%
 Department of Physics, Hiroshima University,
 Higashi-Hiroshima 739-8526, Japan
}

\recdate{October 16, 2002}

\abst{
   We have explored the magnetosphere described by 
a numerical solution for an axisymmetric rotating
field satisfying the force-free and ideal 
MHD conditions everywhere.
The electric current distribution is determined
by the requirement of continuity and smoothness 
of the flux function across the light cylinder.
The overall magnetic field structure is obtained 
using the numerical result.
We also checked the drift velocity
of a charged particle, 
which is caused by the resultant electric and 
magnetic fields.
It is found that the velocity exceeds the light speed
beyond a few times the size of the light cylinder.
This fact suggests the breakdown of 
some underlying assumptions.

}

\begin{document}

\date{\today}
\maketitle

\section{Introduction}
It is well known that
rotating magnetized neutron stars emit
radiation from radio waves to gamma rays, and 
constitute a power supply to surrounding nebula.  
The basic idea of a pulsar 
was proposed
more than 30 years ago $\bigl($e.g., Refs.~\citen{gj69} and
\citen{og69}$\bigr)$, 
but
a detailed consistent model has not yet been constructed.
Almost all the energy is carried out 
from a pulsar by electromagnetic fields, but
it is converted to particle kinetic energy
at large distances\cite{rg74}. 
Several authors have discussed the conversion mechanisms,
such as non-ideal MHD effects \cite{ms94}, 
magnetic reconnection \cite{co90,lk01}
and plasma instability \cite{mm96,be98}.
Because the dissipation of electro-magnetic fields
is related to the acceleration of particles, and
hence the observed radiation, 
the determination of the mechanism is very important.
%

The global structure of the magnetosphere, even 
without dissipation, is not easily determined.
The configuration is strongly coupled
with the plasma flow.
Several authors have used  simplified
electromagnetic models, such as a monopole magnetic field,
to investigate the plasma dynamics.
For instance, Michel \cite{mi69,mi73}
considered  cold MHD winds for an aligned pulsar,  and
Bogovalov \cite{bo99} extended it to the oblique case.
%
%
%
The monopole magnetic field itself is unrealistic,
but some results may be used as a model of pulsar magnetosphere.
It is desirable to consider the plasma dynamics
with a more realistic magnetic field, e.g., a dipole field
near the star.
We may gradually confirm the existing general picture, or we may
find incorrect aspects of previous models, 
using a realistic model.
%

%
The pulsar equation determines
an axially symmetric MHD field in the force-free limit
 $\bigl($e.g., Refs. \citen{be97,bgibk,mibk}$\bigr)$.
Recently,
Contopoulos et al. \cite{ckf99}
obtained a new type of numerical solution of
the  pulsar equation.
We refer to their solution as the CKF solution in this paper.
Their solution represents 
the interesting structure that 
the magnetic field is a dipole near the star and 
asymptotically approaches a split-monopole form at far distances.
Their results also show that 
the solution is smooth across the
light cylinder, and there is no evidence of a singularity 
of the flow and field.
In a subsequent paper\cite{ck02}, Contopoulos and Kazanas
proposed a possible conversion mechanism of pulsar power into particle 
energy in the wind region, using their solution.

%
For the above reason, the CKF solution is quite interesting. 
In particular, the non-singular property is 
{\it  in stark contrast with
the previous understanding}
that a dissipation-free
flow would lead to a singularity in the flow and 
the magnetic field, occurring
a short distance beyond the light cylinder
$\bigl($e.g., Ref.~\citen{ms94}$\bigr)$.
The drift approximation is expected to 
break down near the light cylinder
$\bigl($e.g., Ref.~\citen{bgibk}$\bigr)$,
since the drift velocity  
$| \vec{v} _\perp |~=~c| \EV \times \BV |/B^2 $ 
approaches the light speed $c$
due to the fact that $| E | \sim | B |$ in the wind region.
These arguments sound reasonable, but depend strongly 
on the electro-magnetic field configuration.
If there were no drawbacks of the CKF solution,
it would allow for 
the understanding of the pulsar magnetosphere to progress significantly,
 or 
it would reveal the need to change the current
picture presented in the literature \cite{ms94,bgibk}.
On the other hand, 
if the validity of the CKF solution could 
be checked, this solution could be used as the lowest 
approximation, and various physical effects
could be added to it.
Furthermore, it would help to understand other
astrophysical phenomena,
because the pulsar equation
can also be applied to the black hole magnetosphere and jets
 $\bigl($e.g., Ref.~\citen{be97}$\bigr)$.
Therefore, it is quite important to determine the properties
of the CKF solution.
%



The purpose of this paper is to explore the physical 
quantities 
described by the CKF solution. 
In particular, we explicitly calculate the electro-magnetic structure 
and drift velocity.
These quantities 
were not given in Ref.~\citen{ckf99}.
We calculated them    by 
differentiating the magnetic flux function.
For the numerical differentiation,  it was necessary  to construct a
flux function using many more grid points than in the computation
reported in Ref.~\citen{ckf99}.
Furthermore, the  validity of the CKF solution is investigated.
%
%
%
%
In \S 2, we briefly summarize the basic equations for
the axially symmetric MHD field in the force-free limit.
In \S 3, we briefly summarize the numerical method and
the CKF solution constructed using numerical calculations.
In \S 4, the magnetic field structure and
drift velocity are calculated. 
Finally,  some remarks are given in \S 5.
In the appendix, we also briefly summarize 
some properties of the split-monopole 
magnetic field, which can be 
calculated analytically, and is useful for 
comparison.
%

%


%

%


%

\section{ Assumptions and formalism}
In this section we briefly summarize
the equations governing the stationary axisymmetric 
magnetosphere.
In the axisymmetric case, it is convenient to decouple 
the poloidal and toroidal parts of the magnetic fields 
 in cylindrical coordinates, $(R, \phi, Z)$.
The poloidal field $\BV_p = (B_R, ~B_Z)$ is written 
in terms of the flux function $\Phi(R,Z)$, while 
the toroidal field is given by another function, $A(R,Z)$, as
\begin{equation}
 ( B_R, ~B_\phi, ~B_Z)
= \left(
- \frac{1}{R} \partial _Z \Phi , 
~~~\frac{1}{R} A,
~~~\frac{1}{R} \partial _R \Phi 
\right) .
\end{equation} 
On the basis of some assumptions, the function $A$ should depend only on 
$\Phi$, as discussed below.
The plasma is assumed to be frozen to the magnetic 
field everywhere.
In the limit of large magnetic Reynolds number, 
the velocity component perpendicular to  the magnetic field
is described by the so-called 
`` $ E \times B$ '' drift velocity,
\begin{equation}
 \vec{v} _\perp  = \frac{ c \EV \times \BV }{B^2 }.
\label{drftvl}
\end{equation}
Including the velocity along the magnetic field line,
the electric and magnetic fields satisfy the relation
\begin{equation}
 \EV + \frac{ \vec{v} }{c} \times \BV =0.
\label{mhd.cnd}
\end{equation}
This condition holds if there exists
a sufficient amount of charge density, whose origin in discussed
in the literature\cite{bgibk,mibk}.   
%

%
As a first step to understanding the overall structure
without a realistic dissipation mechanism,  
we here ignore the small gap regions and
consider the electro-magnetic structure 
on a much larger length scale.
%
From the ideal MHD condition (\ref{mhd.cnd}),
the electric field can be expressed  by
\begin{equation}
( E_R, ~E_\phi,  ~E_Z )
 = 
\left( -\frac{\Omega}{c} \partial_R \Phi,
~~
0,
~~
- \frac{\Omega}{c} \partial_Z \Phi \right)
= - \vec{\nabla} \left(
 \frac{\Omega}{c} \Phi \right),
\label{eqn.efld0}
\end{equation}
where the angular velocity of the magnetic field line  $\Omega$ 
is also a function of $\Phi$ in general, 
but we assume it to be constant 
 in this paper. Therefore, the magnetosphere rotates like a rigid body.

The charge density $ \rho_e $ is given by
the divergence of the electric field (Gauss's law), 
and
the current $\vec{j}$ 
by the rotation of the magnetic field (Ampere's law).
They are explicitly written as
\begin{eqnarray}
 \rho_e &=& 
 -\frac{\Omega}{4 \pi c} \Delta  \Phi
=-\frac{\Omega}{4 \pi c} 
\left(
\frac{1}{R} \partial_R R \partial_R \Phi 
+ \partial_Z ^2  \Phi
\right), \\
 ( j_R, ~j_\phi,  ~j_Z )
&=& \frac{c}{4\pi}
 \left(
 -\frac{1}{R} \partial_Z A ,
~~
 -\frac{1}{R}\left(
\Delta \Phi    -\frac{2}{R} \partial_R \Phi 
\right) ,
~~
 \frac{1}{R}  \partial_R A  
\right) .
\end{eqnarray}

We assume that the electro-magnetic force dominates.
That is, we ignore the gravity, pressure and 
the inertial terms. 
The force-free condition is written 
\begin{equation}
 \rho_e \EV + \frac{\vec{j}}{c}  \times \BV =0.
\end{equation}
After some algebra, 
we find that
the function $A $ is on a surface of constant $\Phi $,
i.e.   
$A = A(\Phi)  $.
The function $ \Phi $ should satisfy the so-called
`pulsar equation',
\begin{equation}
 \left( 1- \beta^2  \right)
 \Delta  \Phi 
 -\frac{2}{R} \partial_R \Phi 
+ A A ' =0,
\label{eqn.mst} 
\end{equation}
where $ \beta $ is the cylindrical radius normalized 
with respect to
the light cylinder:
$ \beta = R \Omega/c =R/R_{LC}.$
Using Eq.~(\ref{eqn.mst}), 
the current and charge density can be rewritten as
\begin{eqnarray}
( j_R, ~j_\phi,  ~j_Z )
&=& \frac{c}{4\pi}
\left(
A 'B_R , 
~~
-\frac{1}{R} 
\frac{2 \beta ^2 B_Z - A A' }
{ 1 - \beta ^2 },
~~
A 'B_Z 
\right) ,
\label{current} \\
\rho_e &=& 
-\frac{\Omega}{4 \pi c} 
\frac{2B_Z - A A' }
{ 1 - \beta ^2 }.
\label{charge}
\end{eqnarray}
%
The current flow within the flux tube $\Phi$ 
is given by
\begin{equation}
J( \Phi )
= \int _0  ^{\Phi   } \vec{j}  \cdot d \vec{S} 
= \int _0  ^{\Phi   } 
\frac{1}{4\pi R} A' | \nabla \Phi | 
\frac{2\pi R d \Phi}{  | \nabla \Phi | } 
= 
\frac{c}{2}  A( \Phi ), \label{totalcurrentA}
\end{equation}
where we have defined $\Phi = 0$ as the $Z$-axis. 
Therefore, 
\begin{equation}
 A(0) =0 \label{boundaryA}
\end{equation}
is obtained
from the regularity of the toroidal magnetic field
(see~\S\ref{sec.Numerical}).

\bigskip 
\section{ Numerical scheme to solve the singular equation } 
\label{sec.Numerical}
\subsection{ Boundary conditions }
The pulsar equation (\ref{eqn.mst}) 
is seen to have a singular surface, i.e. the light cylinder 
$R  = R_{LC}  = c/\Omega $.
Both inside  and outside the light cylinder,
the equation becomes a partial 
differential equation of the elliptic type for a given function $A$.
The procedure for constructing a numerical solution of 
(\ref{eqn.mst}) is to solve the equation  
on each side and to match the solutions
at the light cylinder.
The interior and exterior solutions are not in general
continuous across the light cylinder for an assumed  
function $A$. Therefore, the function $A$ should be 
modified iteratively  so as to converge.

We set boundary conditions for the elliptic equation 
that are almost the same as for the CKF solution.
%
%
The function $\Phi$ near the star is described 
by a dipole magnetic field with 
magnetic dipole moment $m$ as
\begin{equation}
  \Phi =  \frac{m R^2}{(R^2 + Z^2)^{3/2} } .
\end{equation}
The last open field line for this dipole field
corresponds to $ \Phi = \Phi_{pc} \equiv  m/R_{LC} $, 
which passes the light cylinder on the equatorial plane. 
At the polar axis, $\Phi$ is
given by the Dirichlet condition as 
\begin{equation}
  \Phi( 0, Z) = 0.    
\end{equation}
%
The boundary condition at the outer right points
is given by the Neumann condition as
\begin{equation}
   \partial_R \Phi(R_{{\rm max}}, Z) = 0,
\end{equation}
while, the upper outermost boundary is given by 
\begin{equation}
   \Phi(R, Z_{{\rm max}}) = 0,
\end{equation}
where the  domain of calculation is limited to 
$0 \le  R \le R_{{\rm max}}, ~~ 0 \le  Z \le Z_{{\rm max}} $. 
The conditions
$R_{{\rm max}} = \infty$ and 
$Z_{{\rm max}} = \infty$, were applied to the CKF solution, 
using a certain coordinate transformation, 
but we limited the spatial domain as 
$R_{{\rm max}} = Z_{{\rm max}} =  9 R_{LC}$.
The effect of 
this change was numerically checked, and we found 
that there is no substantial difference
between the results obtained in the two cases.
Our choice is better to facilitate numerical convergence.
We also changed the outer boundary conditions, but we found
that they are not so sensitive to the boundary conditions, 
and the calculations converge to the result almost uniquely.
%
The boundary condition on the equatorial plane is divided by 
the light cylinder.
Because particles can rotate with the star within the 
light cylinder,  we have
\begin{equation}
\partial_Z  \Phi(R, 0) = 0. ~~(R\leq R_{LC})
\end{equation}
Outside the light cylinder, the line is extended to infinity.
We therefore set it as  
\begin{equation}
  \Phi(R, 0)  =  \Phi_{{\rm open}}, ~~(R>R_{LC})
\end{equation}
where $ \Phi_{{\rm open}} $ is the last 
inner  open field line,  i.e.
$\Phi(R_{LC}, 0)$ calculated for the interior solution.

In order to avoid the singular surface of
Eq.~(\ref{eqn.mst}), 
we require at the light cylinder 
$R  = R_{LC} = c/\Omega$
the relation
\begin{equation}
A A'  = \frac{2}{R} \partial_R \Phi =2B_Z  .
\end{equation}
In this case, the singularities of  
the charge density $\rho_e$ and toroidal current $j_\phi $ 
vanish simultaneously 
[See Eqs.~(\ref{current})--(\ref{charge}).]

It is not clear {\it a priori} which functional form 
of $A(\Phi)$ satisfies all of the boundary conditions.
We therefore need to carry out an iterative process
to obtain the solution.
Contopoulos et al. \cite{ckf99}
successfully solved the equation 
by searching for a suitable function  $A$
so as to match the interior and exterior
solutions at the surface.
We used the same 
relaxation-type technique.
%
%

\subsection{ Results } 
\label{subsec.Numerical.Results}
%
We solved Eq.~(\ref{eqn.mst}) with the boundary conditions
in the numerical domain 
$R_{{\rm max}} = Z_{{\rm max}} =  9 R_{LC}$.
%
%
A self-consistent solution for  $\Phi$ and $A(\Phi)$ 
was calculated on a grid with
80 $\times$ 80 points inside and 
another 80 $\times$ 80 points outside the light cylinder
to insure numerical accuracy. 
The calculation required more than $10^{4}$
iterations to realize a smooth result.
We performed many trials to search for 
different types of solutions, but just one, the CKF solution,
was found.
This may be a unique solution satisfying the 
boundary conditions, although we have not yet proved this.
The overall structure of the solution is the same as 
that of
the CKF solution, but the value of the 
last open field line is a little bit larger than 
that of the actual CKF result.
The result in that case, obtained with a 30 $\times$ 30 grid, is
$ \Phi _{{\rm open}} = 1.36 \Phi _{pc} $, while
our result is $ \Phi _{{\rm open}} = 1.66 \Phi _{pc}$.
The origin of this discrepancy is presumably the
difference in numerical resolution. 
(We used more than twice as many grid points
in each direction.) We also checked 
their result with a calculation using a 30 $\times$ 30 grid. 
We found that the value $ \Phi _{{\rm open}} $ 
increases slightly with the grid points, but our numerical result seems
to converge to a value  around $ \Phi _{{\rm open}} = 1.66 \Phi _{pc} .$
 It is interesting to compare this value with
that for the case of constant $ A$  \cite{mi73},
in which $ \Phi _{{\rm open}} = 1.592 \Phi _{pc}$.
The precise value depends on the numerical 
accuracy, and scales the physical quantities of the solution.
The calculated function $\Phi(R,Z) $ is shown in Fig.~1.
Both interior and exterior functions connect smoothly 
across the light cylinder.
%
%
We only show the result for the inner part of the numerical calculation,
where
$ 0 \le R/R_{LC} \le 4 ,~ 0 \le Z/R_{LC} \le 3$.
The result in
this region is almost unaffected by our choice of the outermost
boundary conditions at $R_{{\rm max}}$, $Z_{{\rm max}}$.
%
%

The function needed during the numerical iterations
is not $ A$, but $ AA'$.  
The conditions adopted in our numerical calculation are 
$ AA' = 0$ for  $ \Phi = 0$ and
 $ \Phi \geq \Phi _{{\rm open}}$,
 where we used Eq.~(\ref{boundaryA}) and  the
 implicit assumption that the electric circuit is closed 
in the whole system: 
\begin{equation}
 A(\Phi) = 0. ~~(\Phi \geq \Phi _{{\rm open}})
 \label{bndryA}
\end{equation}
%
The  function $ A$ is constructed from the numerical data
for  $ AA'$.
The result is shown in Fig.~2.
The result indeed shows that $ A( \Phi_{{\rm open}}) \ne  0$. 
This causes serious problems when applying our computation 
to  physical situations.
The current closure is not realized, 
and there is a discontinuity of $B_\phi$ in this model.
The total current from the upper hemisphere has a definite sign, $ J  < 0$.
A simple way to use this model is to introduce
 a thin current sheet on 
$\Phi_{{\rm open}}$.
This artificially ensures that
a return current sheet equal to $-A( \Phi_{{\rm open}})$ flows along the
equatorial plane $(R > R_{LC})$ and the last open field line.
The discontinuity of $B_\phi$ on $\Phi_{{\rm open}}$ is clear, because
$B_{\phi} \ne 0$ along the open field line,
and $B_\phi =  A(\Phi)/R = 0$ on the closed line.
This discontinuity may be related to the current sheet.
However, in a model with the  current sheet, an enormously strong force,
 such as that due to a pressure, would be required.
%
%
%

One may think that a stringent condition, such as $A = 0$ at $\Phi = 0$
and $\Phi_{{\rm open}} = 0$, should be imposed
in order to construct a more realistic model 
(without an artificial current sheet).
We used this condition in the numerical iteration, 
but we did not obtain a convergent solution.
Our numerical experiment may not be complete, 
but it suggests that a global solution of the pulsar equation can be
obtained only under the weak conditions on $A$ used in
previous works\cite{ckf99}.

We also compared the functional form of our result with that in the
split-monopole case.
The interesting point regarding the CKF solution is 
the existence of separated regions with 
different signs of $ A'$. 
The poloidal current flows along a constant
$ \Phi  $  surface, and the current in the interval
$ d\Phi  $  is $ dJ = c A' d\Phi /2 $.
The region in which $A' < 0$ $(0 \le \Phi <1.36 \Phi_{pc})$ 
corresponds to an out-flowing electron
(or an in-flowing positron), and
that in which $A' > 0$ $(1.36 \Phi_{pc} < \Phi \le \Phi_{{\rm open}} )$ 
corresponds to an in-flowing electron. 
However, the amount of flow in 
the region satisfying $A' > 0$ in this model is
insufficient to satisfy the current closure condition,
$J=0$.

\begin{figure}[ht] 
 \centerline{
    \includegraphics[bb= 0 0 504 360 ,scale=.7, origin = c, angle = 0]
    {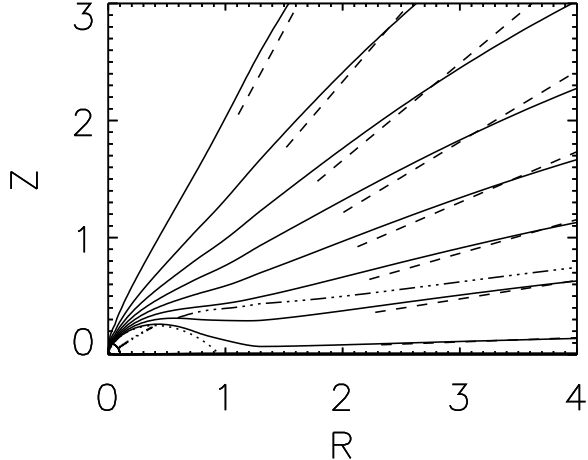}}

 \caption{
 Structure of the axisymmetric force-free magnetosphere of an aligned
  rotating magnetic dipole. The solid curves represent the flux surfaces in
  intervals of 0.2$\Phi_{pc}$, with $\Phi = 0$ along the axis, while
  dashed curves represent those of the split-monopole
  solution.
  The dotted curve represents the last open field line,
   $\Phi_{{\rm open}} = 1.66\Phi_{pc}$.
  The boundary condition near the star is imposed 
  at the quarter circle indicated in the lower left-hand corner of the
  figure.
  The dotted-dash curve indicates the null surface.}  

 \label{psi_with_splitmonopole}
\end{figure} 

\begin{figure}[ht]  
 \centerline{
  \includegraphics[bb= -1 -1 284 406 ,scale=.7, origin = c, angle = 0]
  {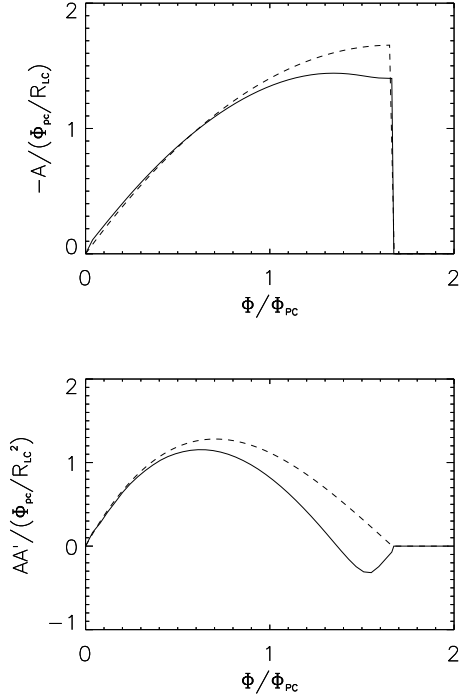}  }
 \caption{The electric current distribution along open field
  lines.
 The solid curve represents our numerical results. It  is similar to
 the CKF solution, but is stretched to larger values of $\Phi_{pc}$.
 The dotted curve represents the results for the split-monopole solution.
  }
  \label{graph_of_A}  
\end{figure}

\section{ Properties of the CKF solution }
\label{sec.Prop}
\subsection{ Global magnetic field structure }
\label{sbsec.Prop.Global}
In Fig.~3, we display the magnetic fields derived 
from the numerical solution. 
The field strength of each component is 
represented by the contours. 
We also show the magnetic fields 
for the split-monopole solution for comparison.
One obviously different region is 
inside the last closed line.
The overall magnetic field structure is
almost the same, except in certain regions.
This seems reasonable, since
the difference between the fields in the 
functions $\Phi$ and $A$ is not so large,
as shown in \S 3.

%
We now consider the differences in detail.
For $ B_R $ of the CKF solution, 
%
there exists a discontinuity on the equatorial plane  around $(R_{LC}, 0)$.
This behavior results from the boundary condition
on the equatorial plane.
As described in  \S 3, the numerical boundary conditions
are divided by the point $(R_{LC}, 0)$.
The condition inside the light cylinder
ensures $ B_R =0$ on the equatorial plane
and $ B_R \ne 0$  in general outside the light cylinder.
Therefore, it might be a subtle problem to devise a proper
numerical treatment around the point $(R_{LC}, 0)$.

Constructing numerical treatment around the $Z$-axis may also be subtle.
In the numerical calculation,
we imposed the condition $ A \to 0$ as $ \Phi \to 0$ (i.e. as $ R \to 0$),
but we have to determine
the value $B_\phi =A/R$ for $R \approx 0$.
The regular field $B_\phi $ should approach  0.
In our numerical calculation, we do not impose
such a strong condition as
$A/R \to 0$.
Therefore, there may be  cusps on the $Z$-axis. 
Similarly, the $B_Z$  field may be singular
on the $Z$-axis. 
We set the boundary condition as 
$\Phi =0$ on the axis and numerically confirmed that
$\partial_R \Phi \approx 0$, but there is no need for
the smoothness of the numerical 
value $B_Z=\partial_R \Phi /R $
near  $R = 0$.
Our numerical results show 
that there is
possibly singular behavior of 
physical quantities,  
$B_R$ near $Z = 0$, and
$B_\phi$ and  $B_Z$ near $R \approx 0$, 
but
we need more precise calculations to
reach a definite conclusion.
%

In Fig.~4, 
we show the ratio of the toroidal to poloidal
components of  the magnetic field, 
$B_\phi/ (B_R ^2 + B_Z ^2 )^{1/2} \equiv B_{t}/ B_{p} $.
Regarding the split-monopole field,
the toroidal component gradually increases with
$R$ and dominates completely for $ R > R_{LC}$.
It also can be seen from our result that
the toroidal field dominates,
but not as strongly,
in the exterior of 
the light cylinder.


\begin{figure}[ht] 
 \begin{center} 
  \includegraphics[bb= -1 -1 568 608 ,scale=.6, origin = c, angle = 0]  
  {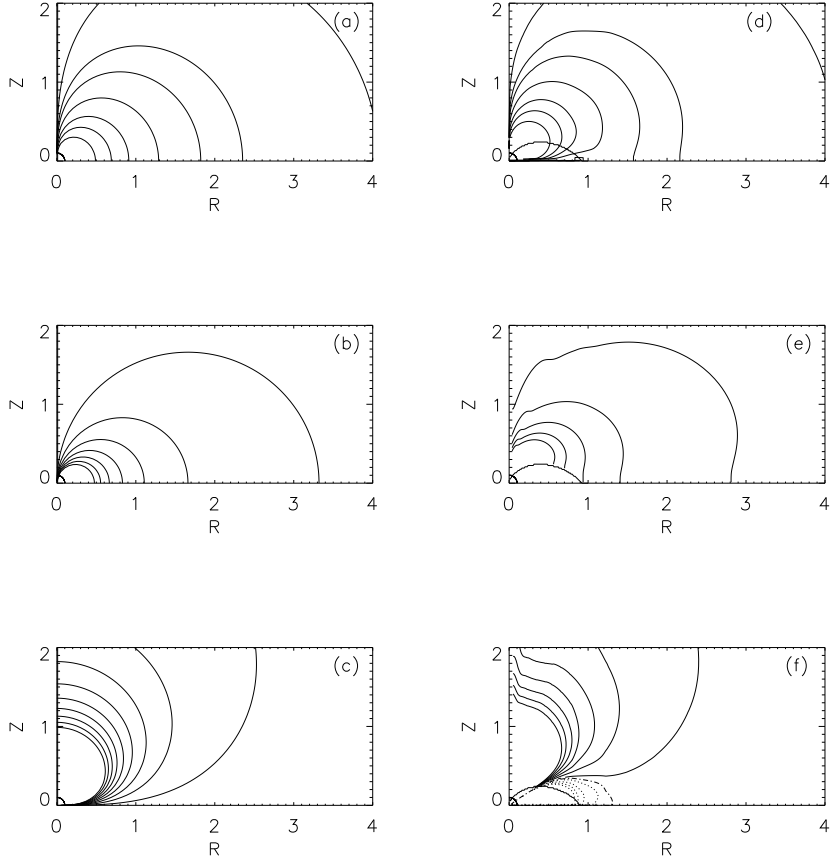}
 \caption{
  Contours of the magnetic field strength for each component 
  of the split-monopole field
  (a)-(c) and our numerical results (d)-(f).
  (a) The solid curves correspond to 
  $B_R$ = 7.0, 3.5, 2.0, 1.0, 0.5, 0.3 and 0.1$B_{o} \equiv m/R_{LC}^{3}$.
  (b) The solid curves correspond to 
  $B_{\phi}$ = $-$3.5, $-$3.0, $-$2.5, $-$2.0, $-$1.5, $-$1.0 and $-$0.5$B_{o}$.
  (c) The solid curves correspond to 
  $B_Z$ = 1.7, 1.5, 1.3, 1.1, 0.9, 0.7, 0.5, 0.3 and 0.1$B_{o}$.
  (d) The solid curves correspond to 
  $B_R$ = 7.0, 3.5, 2.0, 1.0, 0.5, 0.3 and 0.1$B_{o}$.
  (e) The solid curves correspond to 
  $B_{\phi}$ = $-$2.5, $-$2.0, $-$1.5, $-$1.0 and $-$0.5$B_{o}$.
  (f) The solid curves correspond to 
  $B_Z$ = 1.3, 1.1, 0.9, 0.7, 0.5, 0.3 and 0.1$B_{o}$, 
  while dotted curves show outwardly
  $B_Z$ = $-$0.7, $-$0.5, $-$0.3 and $-$0.1 $B_{o}$. 
  The null line, along which $B_Z$ = 0, is indicated by the dotted-dash curve.
 }
  \label{graph of Bph}  
 \end{center}  
\end{figure}

\begin{figure}[ht]  
 \centerline{
  \includegraphics[bb= -1 -1 568 203 ,scale=.7, origin = c, angle = 0]
  {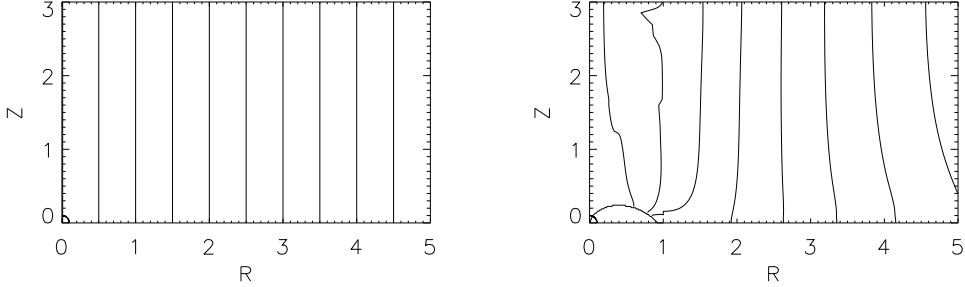}  }
 \caption{
  Contours of the
  ratio of the toroidal to poloidal components of the
  magnetic field.
   The split-monopole solution is shown  in the left panel and our
   result is shown in the right panel.
  The contour level increases from $B_t/B_p$ = 0.5 (innermost)
  at intervals of 0.5.
     The magnitude of the toroidal component 
       exceeds that of the poloidal component beyond the 
light cylinder in both solutions.
  }

  \label{graph of BtBprto}  
\end{figure}

\subsection{ Drift velocity}
The components of the drift velocity (\ref{drftvl})
are explicitly given by
\begin{equation}
  \vec{v} _{\perp} = 
\left(
- \frac{R\Omega }{B^2} B_R B_\phi,~~
 \frac{R\Omega }{B^2} (B_R^2 +  B_Z^2),~~
- \frac{R\Omega }{B^2} B_Z B_\phi 
\right).
\end{equation}
The absolute magnitude is 
\begin{equation}
  v_{\perp} =  R\Omega 
\frac{ (B_R^2 +  B_Z^2)^{1/2} }
{ (B_R^2 +  B_\phi^2 +B_Z^2)^{1/2} } 
= 
R\Omega 
\frac{ B_p }
{ (B_t^2 +  B_p^2)^{1/2} } 
.
\label{vel.lt.c}
\end{equation}
It is clear that the condition $v_{\perp}  \le  c $
holds inside the light cylinder.
However, this condition is not guaranteed outside
the light cylinder, and 
whether or not it holds depends significantly on the
magnetic field configuration.
Indeed, this condition holds everywhere
for the magnetic field of
the split-monopole, as shown in the appendix.
In Fig.~5, we show the drift velocity $v_{\perp} $
calculated from the CKF solution.
The numerical result shows that $v_{\perp}  < 0.8 c$
within the light cylinder  and
that the value $v_{\perp}  $ gradually increases
with $R$.
The drift velocity 
 exceeds the light speed
 for $ R \gtrsim 3 R_{LC}$. 
This violation is clear, especially for small $Z$.
The condition $v_{\perp}  \le  c $
means that  the toroidal magnetic field
should dominate for
$ B_\phi ^2/(B_R^2 +  B_Z^2) \ge (R/R_{LC} )^2 -1 $,
from Eq.~(\ref{vel.lt.c}).
The ratio of the toroidal to poloidal components,
$ B_\phi /(B_R^2 +  B_Z^2)^{1/2} $,
increases with $R$, but the dependence is not as steep
as for the split-monopole field.  
This difference leads to the violation of causality, i.e.
$v_{\perp}  \ge  c $ in the CKF solution,
but no violation in the split-monopole case.
It is thus seen that
the applicable range of the CKF solution
is limited  within  a few times the size of the light cylinder.

\begin{figure}[ht] 
 \begin{center} 
  \includegraphics[bb= -1 -1 568 203 ,scale=.7, origin = c, angle = 0]  
  {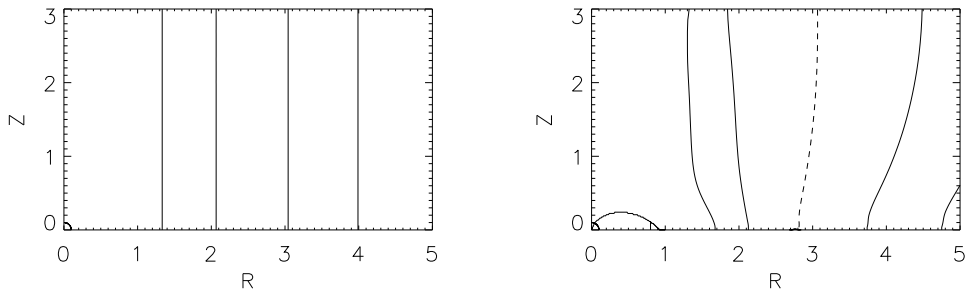}
 \caption{
  Contours of the drift velocity for the split-monopole solution
   and our numerical
  results.
  The solid lines in the left panel indicate 
  $\left|\vec{v} _\perp\right|$ =  0.8,
  0.9, 0.95 and  0.97 $c$. 
  The drift velocity of our result is shown in right-hand 
  panel for the cases
   $\left|\vec{v} _\perp\right|$ = 0.8, 0.9, 1.1 and 1.2$c$
  (solid), and  $\left|\vec{v} _\perp\right|$ = 1.0$c$ (dotted).  }
    \label{Drift V (total)}
 \end{center}  
\end{figure}

   One may expect the breakdown of the drift approximation,
as suggested in the literature\cite{bgibk}, but
it should be noted that 
the breakdown 
depends significantly on the adopted electro-magnetic fields.
A counterexample is the split-monopole case.
In this case, the drift approximation holds everywhere.
The breakdown of the drift approximation is likely to occur
outside the light cylinder for most electro-magnetic fields.
%
In general, the location of the breakdown is not known 
{\it a priori}.
The drift velocity cannot exceed the light 
velocity in a realistic situation. 
The inertial term gradually prevents the increase of the 
drift velocity as a particle moves away from a star.
The magnetic field near a star is strong enough that
the inertial term of the particle can be ignored, but it
decreases with the distance from the star. 
For this reason, the inertial term 
can no longer be ignored beyond a certain distance.
This is a general picture, but the position 
beyond which the inertial term cannot be ignored 
depends strongly on the magnetic field configuration. 
In this paper, we have for the first time
elucidated the properties of electro-magnetic fields 
described by the CKF solution and
the drawbacks of the solution.

\section{ Remarks }

In this paper, we have explored the axially symmetric 
magnetosphere around an aligned rotating dipole field.
The ideal MHD and force-free conditions are satisfied 
everywhere by assuming a sufficient charge density.
The electro-magnetic structure can be determined using the 
so-called pulsar equation, which
is a coupled equation for $\Phi$ and $A$. 
Some authors have examined some special cases in the past,
 e.g.  
that in which 
$A$  is  constant \cite{mi73a}, linear in
$\Phi$ \cite{sw73}, and  quadratic in 
$\Phi$ \cite{mi69}.
However, these do not correspond to realistic situations.
A new numerical solution reported by Contopoulos et al.
\cite{ckf99} seems to be promising.
That solution may be useful as a starting
point for a more realistic pulsar magnetosphere.
From this point of view, we have examined the magnetic field
and drift velocity in detail.  

%
The overall global structure of the magnetosphere
we have studied, is quite similar to
that of a split-monopole field. 
Some difference in the current distribution $ A$ 
appears in regions corresponding to the case in which the field lines
extend above the equatorial plane.
A modification is necessary to connect
the inner dipole-like field with the outer monopole-like field.
This causes return currents, although 
the total amount of such currents is insufficient to realize current closure.
For this reason 
a current sheet is required in the magnetosphere.
%
Our result also shows
the breakdown of the drift approximation 
beyond a few times the size of the light cylinder,
%
%
and
this result may modify the understanding regarding the CKF solution 
proposed in Ref.~\citen{ck02}.
In that work, it was shown that
the Lorentz factor of the outflowing plasma increases 
linearly with the distance from a neutron star.
Therefore, 
at a large distance, $R \sim 10^3 R_{LC}$, the inertial term can no longer
be ignored (i.e., the force-free condition breaks down),
whereas 
our numerical calculation shows that the validity of
the CKF solution is limited within a short distance $\sim 3 R_{LC}$.
Therefore some assumptions used in this calculation should incorporate
the effects of a non-ideal MHD
 or the inertia of the particles in order to obtain  a
more realistic solution $\bigl($e.g., see Ref.~\citen{ms94}$\bigr)$.

\section*{Acknowledgements}
This work was supported in part
by a Grant-in-Aid for Scientific Research 
(No.~14047215) from 
the Japanese Ministry of Education, Culture, Sports,
Science and Technology.

\section*{ Appendix: Split-monopole solution }
In this appendix, we consider the physical 
structure represented by the split-monopole solution. 
It may be useful to compare this explanation 
with the CKF solution. 
As noted by CKF, their solution asymptotically 
approaches the split-monopole solution. 

When the function $A$ is quadric in $\Phi$,
the resultant magnetic field corresponds to 
the magnetic monopole solution. 
The solution of Eq.~(\ref{eqn.mst})
can be explicitly written as 
\begin{eqnarray} 
 \Phi &=& q \left(1-  \frac{Z}{ (R^2 +Z^2)^{1/2} }
\right), \\
 A &=& - \frac{\Omega \Phi }{c}
\left( 2 -\frac{\Phi}{q} \right) ,
\end{eqnarray}
where  $q $ is related to the monopole charge.
This fact is easily checked by calculating 
the magnetic field given as
\begin{equation} 
( B_R, ~B_\phi,  ~B_Z )
 = 
\left(  \frac{ q R}{ (R^2 +Z^2)^{3/2} },
~~
-\frac{ q \Omega R}{c (R^2 +Z^2) },
~~
 \frac{ q Z}{ (R^2 +Z^2)^{3/2} }
 \right).
 \end{equation}
The poloidal component of magnetic field 
$( B_R,   ~B_Z )$ is always radial. 
This poloidal field dominates near the origin.
By contrast,
the toroidal field dominates across the
light cylinder; that is,
$ B_\phi  > B_p =( B_R ^2 +B_Z ^2 ) ^{1/2}$ 
for $R > c/\Omega $.
In this way, this monopole solution represents a
smooth transition 
of the magnetic field through the light cylinder. 
It may be helpful in obtaining  physical understanding 
to calculate the current flow, 
\begin{equation} 
( j_R, ~j_\phi,  ~j_Z )
 = \frac{ q \Omega }{2 \pi c  }
\left(  \frac{  RZ}{ (R^2 +Z^2)^{2} },
~~
0,
~~
\frac{  Z^2 }{ (R^2 +Z^2)^{2} } 
 \right).
\end{equation}
The current flow is purely poloidal and out-flowing, i.e.
$ J>0 $.
This current gradually enhances the toroidal
magnetic field $ B_\phi$, which dominates
outside the light cylinder.

The drift velocity is given by
\begin{equation} 
( v_R, ~v_\phi,  ~v_Z )
 = 
\frac{ c \beta}{  (1 +\beta^2 ) }
\left(  \frac{  R \beta}{ (R^2 +Z^2)^{1/2} },
~~
1,
~~
\frac{  Z \beta}{  (R^2 +Z^2)^{1/2} } 
 \right).
 \end{equation}
It is easily verified that $v_\perp \le 1$ everywhere. 
This drift velocity asymptotically approaches the light speed
as $ R \to \infty $.

%
On the basis of the above consideration, there seems to 
be no problem with the monopole solution,
except that it may not exist in nature.
No monopole has yet been discovered in nature,
and so 
this solution is used only for, e.g., upper hemisphere
with $+q $ 
and the lower hemisphere with $-q $.
Combining these two solutions with opposite charge,  
the so-called split-monopole  solution is sometimes 
used for physical analysis,
e.g., in Ref.~\citen{mi73}.
In that case,
the particle velocity of consistent flow 
is $c$ everywhere.

%


\end{document}